\documentstyle[12pt,epsf,epsfig,wrapfig]{article}
\begin{document}

\begin{center}
{\bfseries PROGRESS IN THE DETERMINATION OF POLARIZED PDFs\\[2mm]
AND HIGHER TWIST}

\vskip 5mm E. Leader$^{1}$, \underline{A.V. Sidorov}$^{2\dag}$ and
D.B. Stamenov$^{3}$

\vskip 5mm {\small (1) {\it Imperial College London, Prince
Consort Road, London SW7 2BW, England
}\\
(2) {\it Bogoliubov Theoretical Laboratory, JINR, Dubna, Russia
}\\
(3) {\it Institute for Nuclear Research and Nuclear Energy,\\
Bulgarian Academy of Sciences, Sofia, Bulgaria }
\\

$\dag$ {\it E-mail: sidorov@theor.jinr.ru }}

\end{center}

\vskip 5mm
\begin{abstract}
The impact of the recent very precise CLAS and COMPASS $g_1/F_1$
data on polarized parton densities and higher twist effects is
discussed. It is demonstrated that the low $Q^2$ CLAS data improve
essentially our knowledge of higher twist corrections to the spin
structure function $g_1$, while the large $Q^2$ COMPASS data
influence mainly the strange quark and gluon polarizations. It is
also shown that the uncertainties in the determination of the
polarized parton densities are significantly reduced. We find also
that the present inclusive DIS data cannot rule out a negative
polarized and changing in sign gluon densities. The present status
of the proton spin sum rule is discussed.

\end{abstract}

\vskip 8mm
\section{Introduction}

One of the features of polarized DIS is that a lot of the present
data are in the preasymptotic region ($Q^2 \sim 1-5~\rm
GeV^2,~4~\rm GeV^2 < W^2 < 10~\rm GeV^2$). This is especially the
case for the experiments performed at the Jefferson Laboratory. As
was shown in \cite{LSS_HT}, to confront correctly the QCD
predictions to the experimental data including the preasymptotic
region, the {\it non-perturbative} higher twist (powers in
$1/Q^2$) corrections to the nucleon spin structure functions have
to be taken into account too.

In this talk we discuss the impact of the recent very precise CLAS
\cite{CLAS06} and COMPASS \cite{COMPASS06} inclusive polarized DIS
data on the determination of both the longitudinal polarized
parton densities (PDFs) in the nucleon and the higher twist (HT)
effects. These experiments give important information about the
nucleon structure in quite different kinematic regions. While the
CLAS data entirely belong to the preasymptotic region and as one
can expect they should mainly influence the higher twist effects,
the COMPASS data on the spin asymmetry $A_1^d$ are large $Q^2$
data and they should affect mainly the polarized parton densities.
In addition, due to COMPASS measurements we have for the first
time accurate data at small $x~(0.004 < x < 0.015)$, which allow
to determine the behavior of the PDFs at small $x$ region and
therefore to calculate more precisely the first moment of the
nucleon spin structure $g_1$.

\section{NLO QCD analysis of the data}

The method used to extract simultaneously the polarized parton
densities and higher twist corrections to the spin-dependent
nucleon structure function $g_1$ is described in \cite{LSS_HT}.
According to this method, the $g_1/F_1$ and $A_1(\approx g_1/F_1)$
data have been fitted using the experimental data for the
unpolarized structure function $F_1(x,Q^2)$
\begin{equation}
\left[{g_1(x,Q^2)\over F_1(x, Q^2)}\right]_{exp}~\Leftrightarrow~
{{g_1(x,Q^2)_{\rm LT}+h(x)/Q^2}\over F_1(x,Q^2)_{exp}}~.
\label{g1F2Rht}
\end{equation}
As usual, $F_1$ is replaced by its expression in terms of the
usually extracted from unpolarized DIS experiments $F_2$ and $R$
and the phenomenological parametrizations of the experimental data
for $F_2(x,Q^2)$ \cite{NMC} and the ratio $R(x,Q^2)$ of the
longitudinal to transverse $\gamma N$ cross-sections \cite{R1998}
are used. Note that such a procedure is equivalent to a fit to
$(g_1)_{exp}$, but it is more precise than the fit to the $g_1$
data themselves actually presented by the experimental groups
because here the $g_1$ data are extracted in the same way for all
of the data sets.

In Eq. (\ref{g1F2Rht}) "LT" denotes the leading twist contribution
to $g_1$
\begin{equation}
g_1(x, Q^2)_{\rm LT}= g_1(x, Q^2)_{\rm pQCD} + h^{\rm TMC}(x,
Q^2)/Q^2 + {\cal O}(M^4/Q^4)~, \label{g1LT}
\end{equation}
where $g_1(x, Q^2)_{\rm pQCD}$ is the well known (logarithmic in
$Q^2$) NLO pQCD contribution
\begin{equation}
g_1(x,Q^2)_{\rm pQCD}={1\over 2}\sum _{q} ^{N_f}e_{q}^2 [(\Delta q
+\Delta\bar{q})\otimes (1 + {\alpha_s(Q^2)\over 2\pi}\delta C_q)
+{\alpha_s(Q^2)\over 2\pi}\Delta G\otimes {\delta C_G\over N_f}],
\label{g1partons}
\end{equation}
and $h^{\rm TMC}(x, Q^2)$ are the calculable kinematic target mass
corrections \cite{TMC}, which effectively belong to the LT term.
In Eq. (\ref{g1partons}), $\Delta q(x,Q^2), \Delta\bar{q}(x,Q^2)$
and $\Delta G(x,Q^2)$ are quark, anti-quark and gluon polarized
densities in the proton, which evolve in $Q^2$ according to the
spin-dependent NLO DGLAP equations. $\delta C(x)_{q,G}$ are the
NLO spin-dependent Wilson coefficient functions and the symbol
$\otimes$ denotes the usual convolution in Bjorken $x$ space. $\rm
N_f$ is the number of active flavors ($\rm N_f=3$ in our
analysis). $h(x)/Q^2$ in Eq. (\ref{g1F2Rht}) corresponds to the
first term in the $(\Lambda^2_{\rm QCD}/Q^2)^n$ expansion of
higher twist contribution to $g_1$. Its logarithmic $Q^2$
dependence, which is not known in QCD, is neglected. Compared to
the principal $1/Q^2$ dependence it is expected to be small and
the accuracy of the present data does not allow its determination.
Therefore, the extracted from the data values of $h(x)$ correspond
to the mean $Q^2$ for each $x$-bean.

Let us discuss now how inclusion of the CLAS EG1 proton and
deuteron $g_1/F_1$ data \cite{CLAS06} and the {\it new} COMPASS
data on $A_1^d$ \cite{COMPASS06} influence our previous results
\cite{LSS05} on polarized PDFs and higher twist obtained from the
NLO QCD fit to the world data \cite{world}, before the CLAS and
the latest COMPASS data were available.

\section{Impact of the new data on polarized PDFs and HT}

The new CLAS $\rm EG1/p,d$ data on $g_1/F_1$ (633 experimental
points) \cite{CLAS06} and the recent COMPASS data on the
longitudinal asymmetry $A_1^d$ (15 experimental points)
\cite{COMPASS06} are at very different kinematic regions. While
the CLAS data are high-precision data at low $Q^2$:$\{x\sim
0.1-0.6,~Q^2\sim 1-5~\rm GeV^2,~W>2~\rm GeV\}$, the COMPASS data
are mainly at large $Q^2$:$\{0.0046 \leq x \leq 0.57,~Q^2\sim
1-55~\rm GeV^2\}$ and the only precise data covering the low $x$
region. Therefore, they will play a different role in the
improvement of the determination of the polarized PDFs and higher
twist effects. The new PDFs and HT and their uncertainties will be
compared with those of LSS'05 determined from our previous
analysis of the world data \cite{world} available before the CLAS
$\rm EG_1/p,d$ and COMPASS'06 data have appeared.

\begin{wrapfigure}[21]{R}{6cm}
\begin{center}
\mbox{\epsfig{figure=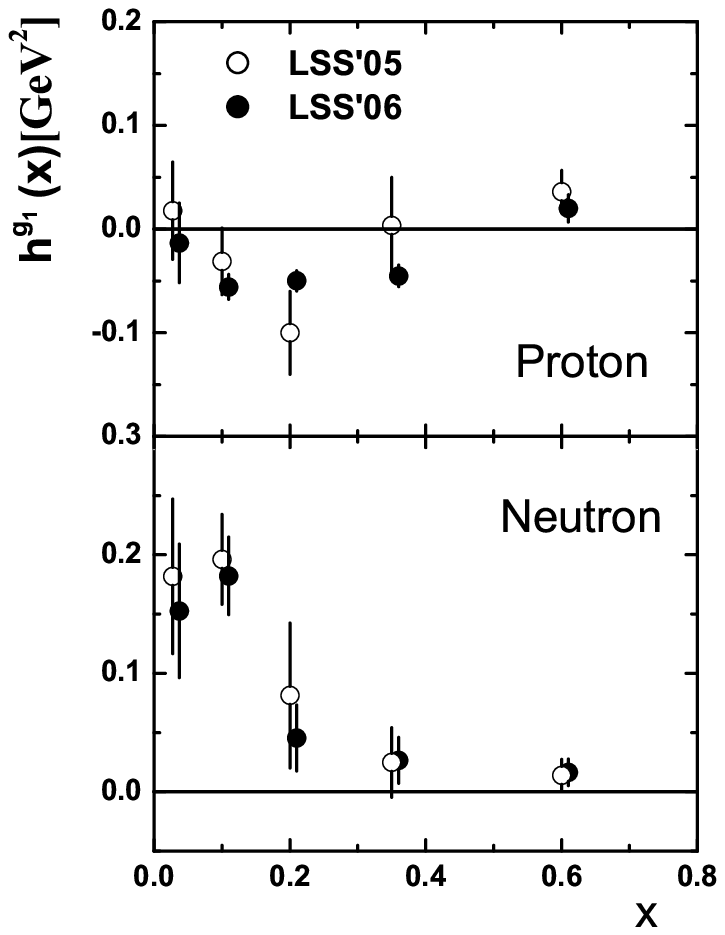,width=6cm,height=7cm}}
\end{center}
{\small{\bf Figure 1.} Effect of new data on the higher twist
values.}
%\label{fig1}
\end{wrapfigure}

As the CLAS data are mainly low $Q^2$ data where the role of HT
becomes important, they should help to fix better the higher twist
effects. Indeed, due to the CLAS data, the determination of HT
corrections to the proton and neutron spin structure functions,
$h^p(x)$ and $h^n(x)$, is significantly improved in the CLAS $x$
region, compared to the values of HT obtained from our LSS'05
analysis \cite{LSS05} in which a NLO(${\rm \overline {MS}}$) QCD
approximation for $g_1(x,Q^2)_{\rm LT}$ was used. This effect is
illustrated in Fig. 1. One can conclude now that the HT
corrections for the proton target are definitely different from
zero and negative in the $x$ region: 0.1-0.4. Also, including the
CLAS data in the analysis, the HT corrections for the neutron
target are better determined in the $x$ region: 0.2-0.4. Note that
$h^n(x)$ at $x \sim 0.5$ was already fixed very precisely from the
JLab Hall A data on the ratio $g^{(n)}_1/F^{(n)}_1$. We have found
that the impact of the COMPASS'06 data on the values of higher
twist corrections and their uncertainties is negligible. The only
exception are the central values of HT at small $x$ for both the
proton and the neutron targets which are slightly lower than the
old ones. Note that this is the only region where the COMPASS DIS
events are at small $Q^2$: 1-4 $\rm GeV^2$.
\begin{figure}[t!]
\begin{center}
\begin{tabular}{cc}
\mbox{\epsfig{figure=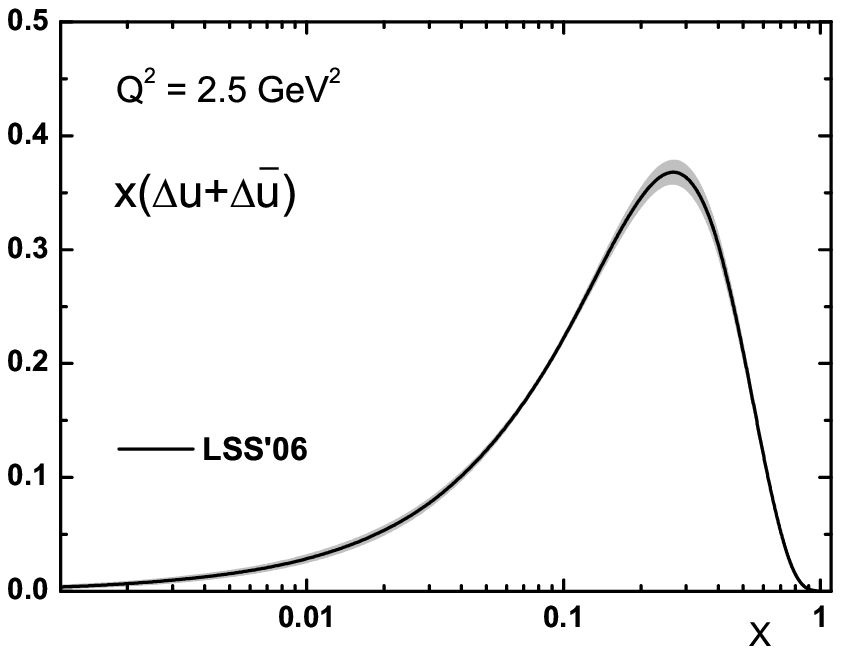,width=6cm,height=4.5cm}}&
\mbox{\epsfig{figure=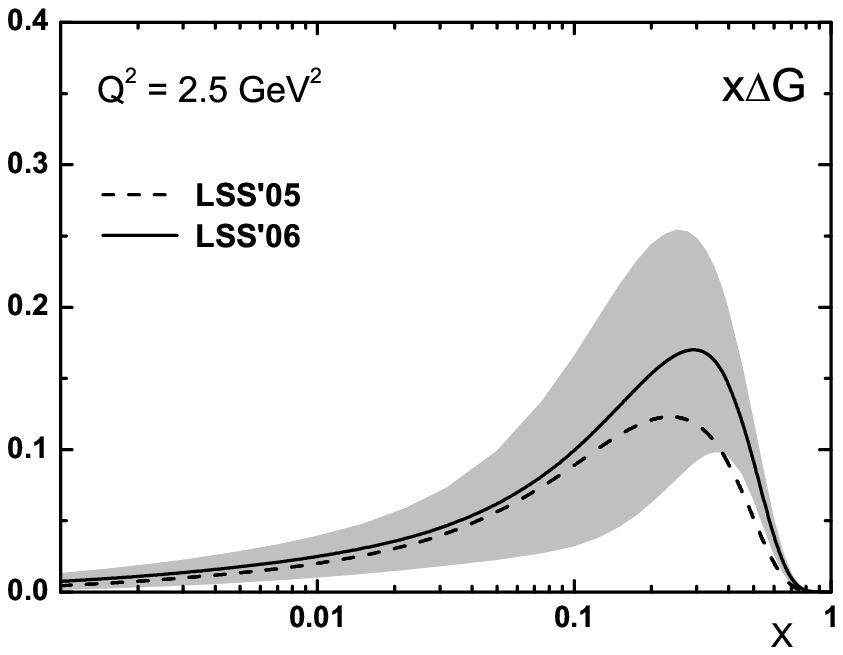,width=6cm,height=4.5cm}}\\
\mbox{\epsfig{figure=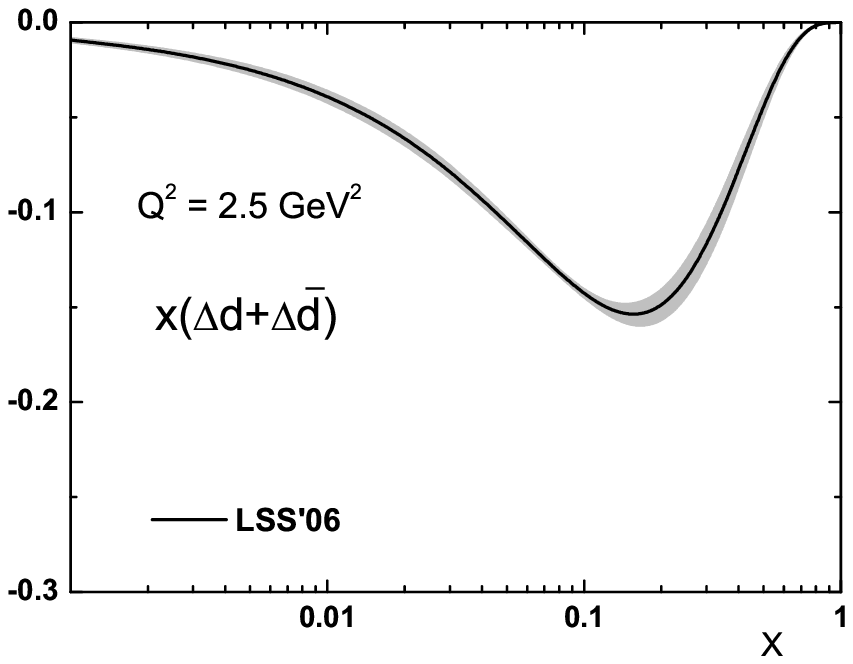,width=6cm,height=4.5cm}}&
\mbox{\epsfig{figure=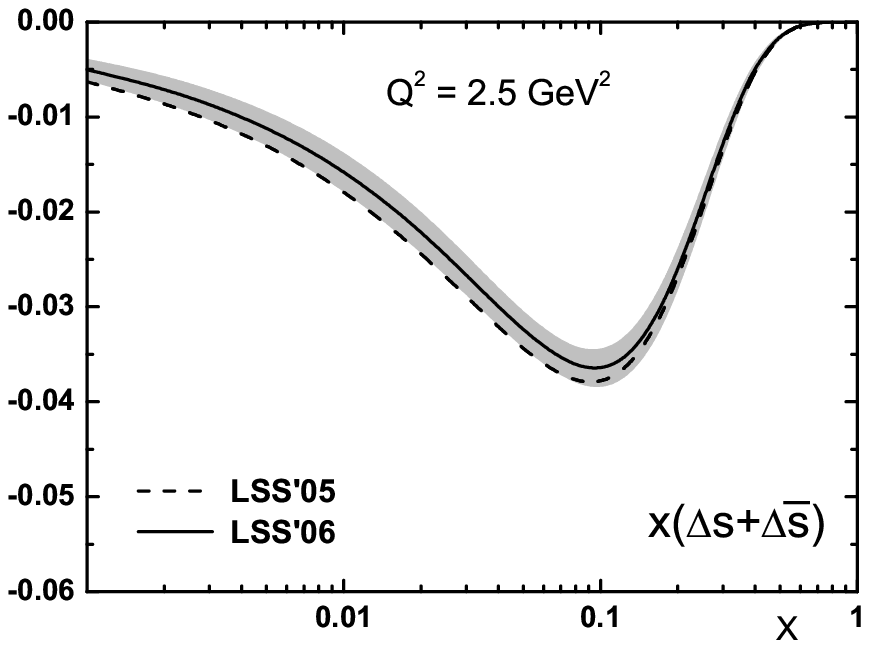,width=6cm,height=4.5cm}}\\
%{\bf(a)}& {\bf(b)}& {\bf(c)}
\end{tabular}
\end{center}
\centerline {\small{\bf Figure 2.} Effect of new data on the
NLO($\rm \overline{MS}$) polarized parton densities. }
\end{figure}
\begin{figure}[tb!]
\begin{center}
\begin{tabular}{cc}
\mbox{\epsfig{figure=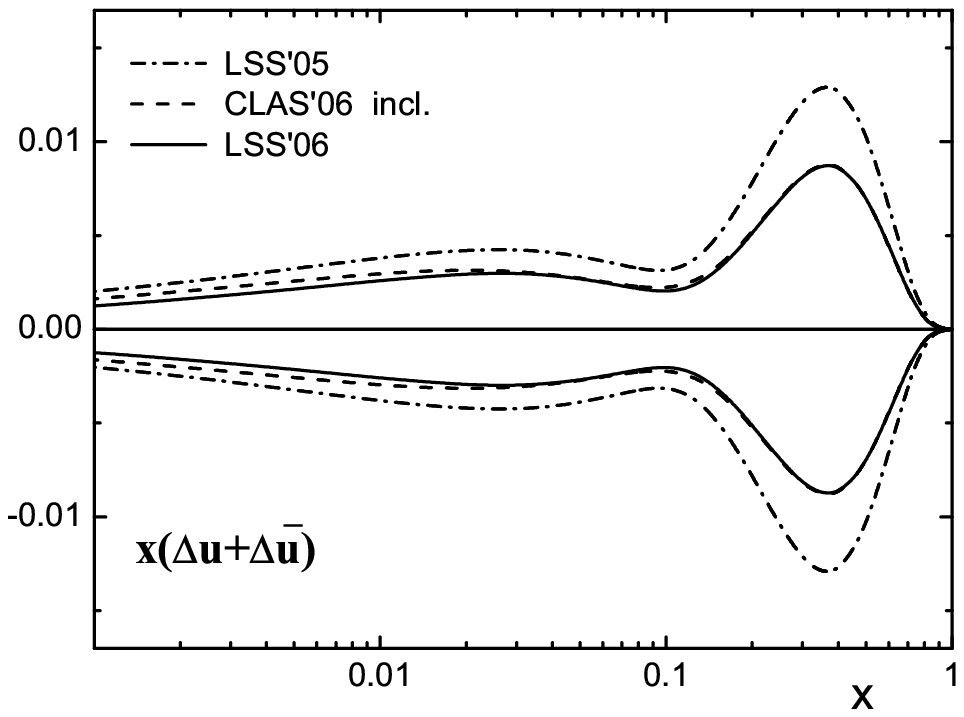,width=6cm,height=5cm}}&
\mbox{\epsfig{figure=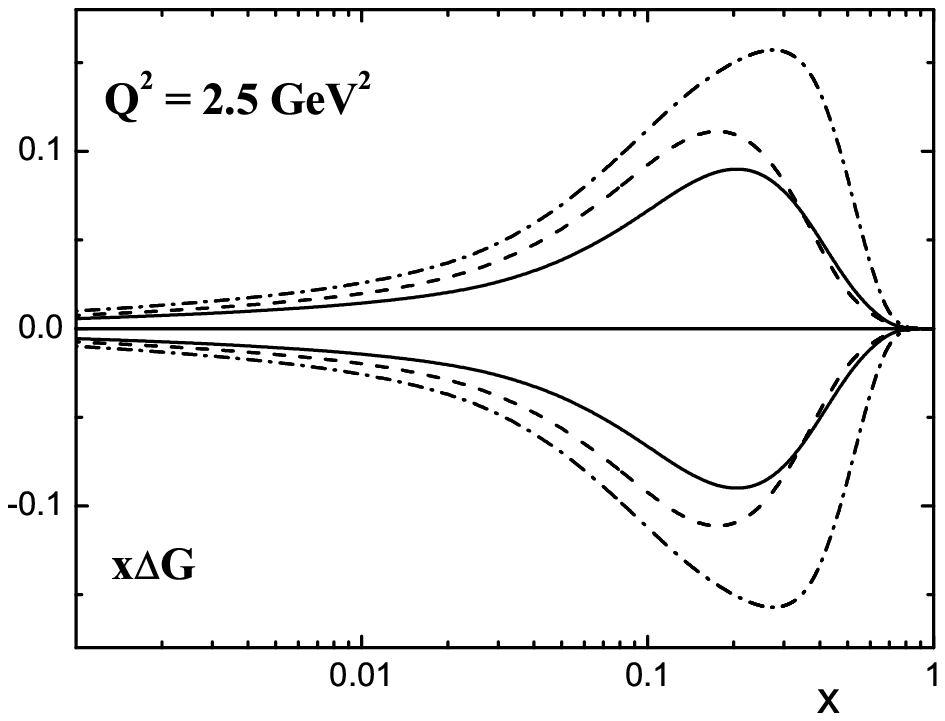,width=6cm,height=5cm}}\\
\mbox{\epsfig{figure=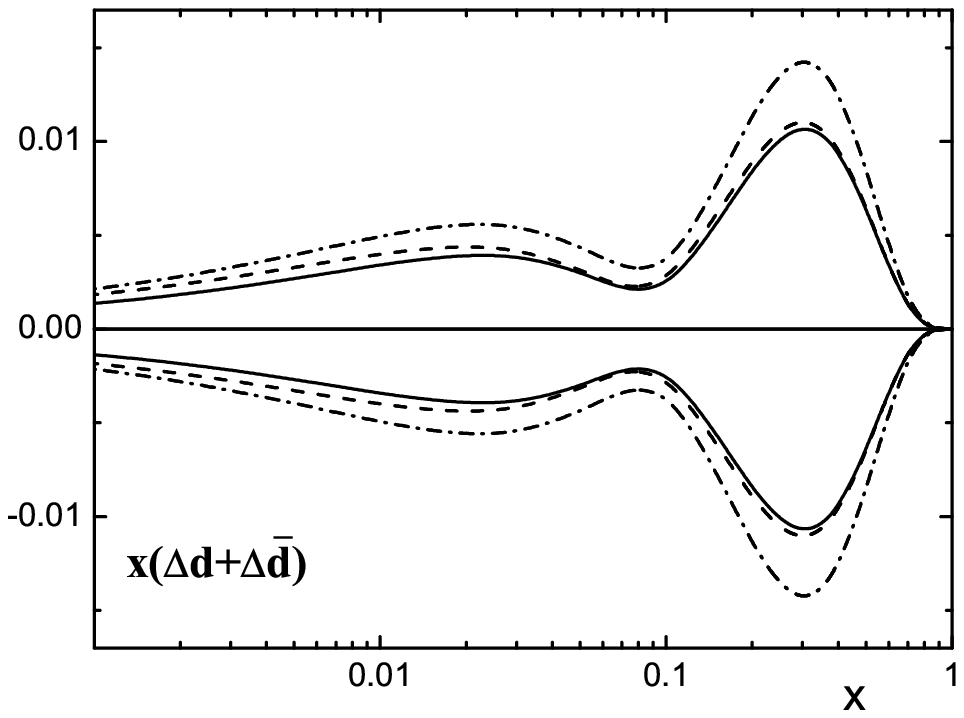,width=6cm,height=5cm}}&
\mbox{\epsfig{figure=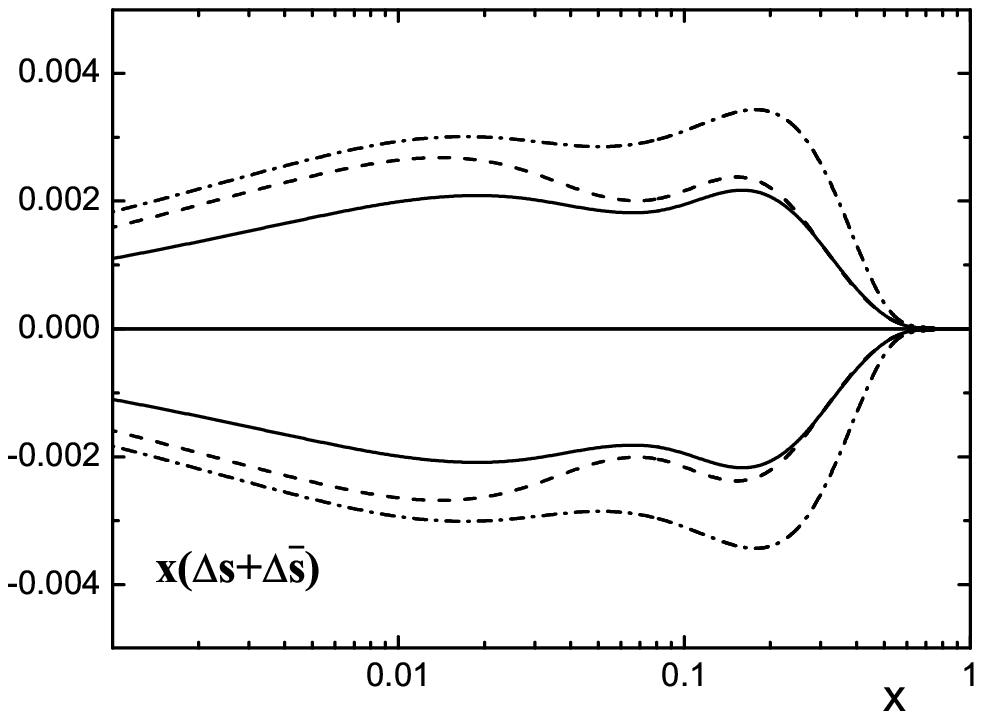,width=6cm,height=5cm}}\\
%{\bf(a)}& {\bf(b)}& {\bf(c)}
\end{tabular}
\end{center}
\centerline {\small{\bf Figure 3.} Impact of GLAS and COMPASS data
on the uncertainties for NLO($\rm \overline{MS}$) polarized PDFs.
}
\end{figure}

The effect of the new data on the polarized PDFs and their
uncertainties is demonstrated in Figures 2 and 3, respectively.
The central values of both the $(\Delta u + \Delta\bar u)$ and
$(\Delta d + \Delta\bar d)$ parton densities do not change in the
experimental region (the corresponding LSS'06 curves can not be
distinguished from those of LSS'05). As one can see from Fig. 2
the new data influence only the polarized gluon and strange quark
sea densities (while the magnitude of strange sea decreases at
$x<0.1$, the gluon density increases at $x>0.1$). As expected, the
central values of the polarized PD are practically {\it not}
affected by the CLAS data. This is a consequence of the fact that
at low $Q^2$ the deviation from logarithmic in $Q^2$ pQCD
behaviour of $g_1$ is accounted for by the higher twist term in
$g_1$ in Eq. (\ref{g1F2Rht}). So, the change of the central values
of the polarized gluon and strange quark sea densities is entirely
due to the new COMPASS data. On the contrary, the accuracy of the
determination of polarized PDFs is essentially improved due to the
CLAS data (the dashed curves in Fig. 3). This improvement is a
consequence of the much better determination of higher twist
contributions to the spin structure function $g_1$, as discussed
above. The impact of COMPASS data on the uncertainties for the
PDFs is also shown in Fig. 3 (the solid curves). As seen, they
help to improve in addition the accuracy of the determination of
the gluon and strange sea quark polarized densities at small $x$:
$x<0.2$ for the gluons and $x<0.1$ for the strange sea.
\begin{figure}[t!]
\begin{center}
\begin{tabular}{cc}
\mbox{\epsfig{figure=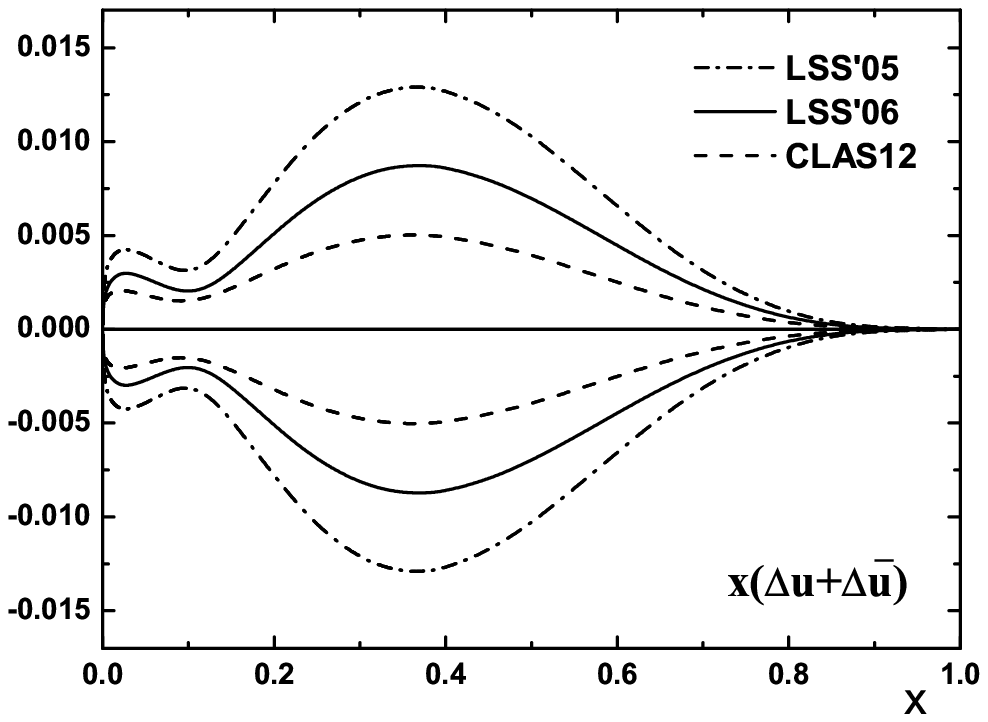,width=6cm,height=4.5cm}}&
\mbox{\epsfig{figure=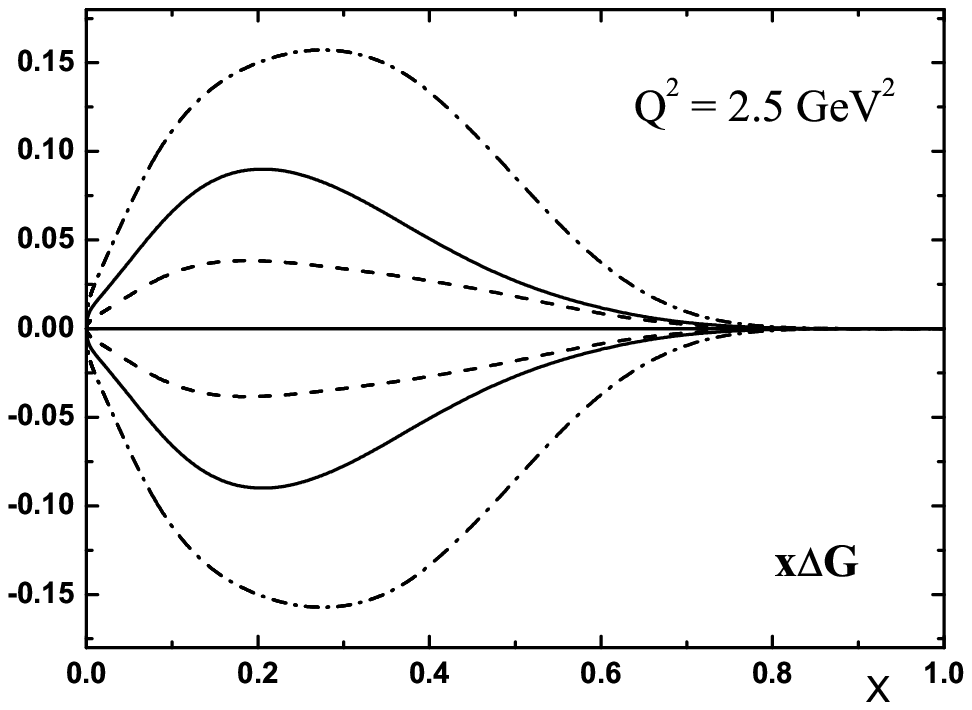,width=6cm,height=4.5cm}}\\
\mbox{\epsfig{figure=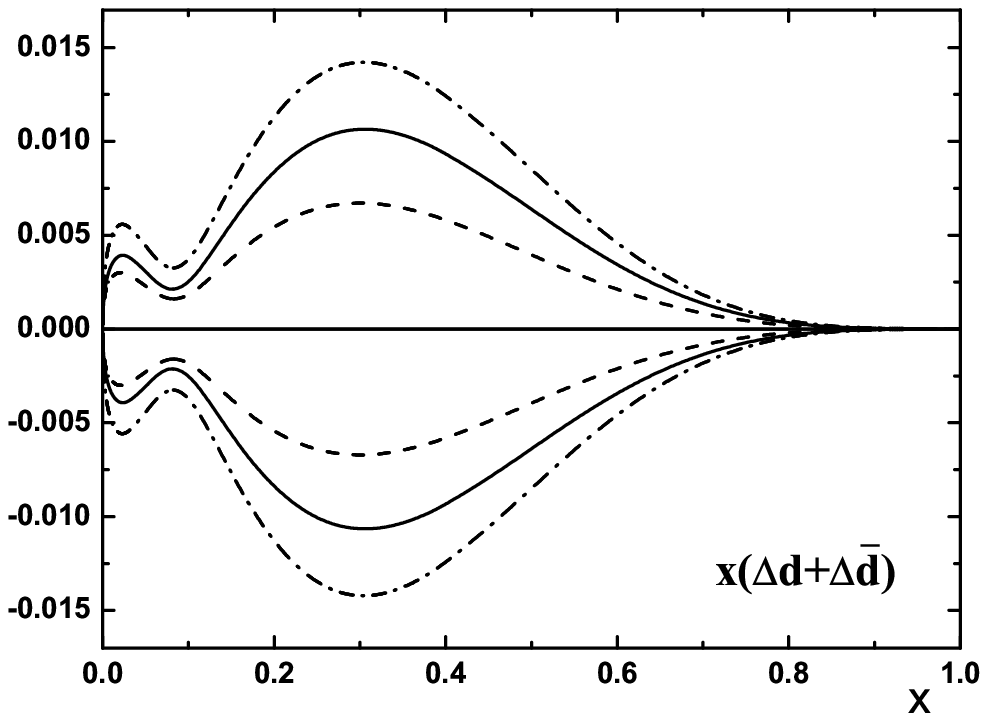,width=6cm,height=4.5cm}}&
\mbox{\epsfig{figure=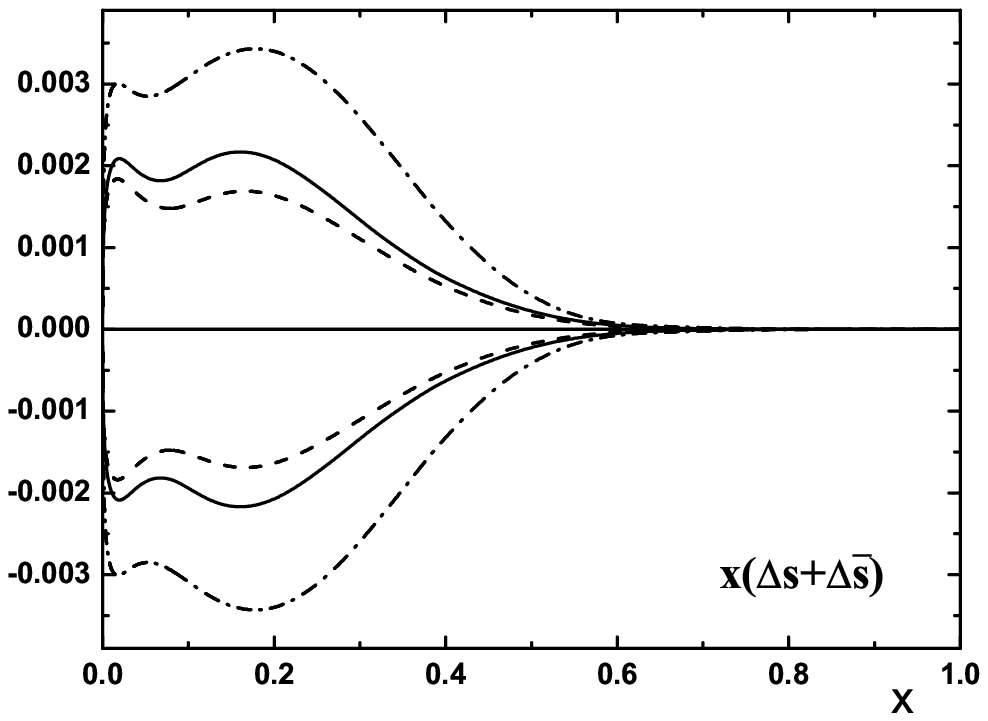,width=6cm,height=4.5cm}}\\
%{\bf(a)}& {\bf(b)}& {\bf(c)}
\end{tabular}
\end{center}
%\centerline
{\small{\bf Figure 4.} Expected uncertainties for
NLO($\rm \overline{MS}$) polarized PDFs after including the data
set to be collected with CLAS12 experiment including statistical
and systematic errors }
\end{figure}

An essential further improvement (the dashed lines in Fig. 4) can
be achieved after including in the analysis the data set to be
collected with CLAS12 experiment \cite{CLAS12} planned to be
performed using a 12 GeV electron beam at Jefferson Laboratory,
USA.

At the end of this Section we would like to mention that all
results on the PDFs presented here have been obtained when 5
$x$-bins have been used to extract the HT values. Due to the good
accuracy of the CLAS data, one can split the measured $x$ region
of the world data set into 7 bins instead of 5, as used up to now,
and therefore, can determine more precisely the $x$-dependence of
the HT corrections to $g_1$. The numerical results of the best fit
to the data using 7 $x$-bins are presented in \cite{LSS06}. It is
important to emphasize that the central values for the PDFs(5
bins) and PDFs(7 bins) excepting the gluons are very close to each
other. However, the uncertainties for the PDFs(5 bins) are smaller
than those for PDFs(7 bins), especially for $\Delta s(x)$ and
$\Delta G(x)$. That is why we prefer to present here the PDFs and
there uncertainties corresponding to 5 bins in $x$ using for the
HT values.

\section{The sign of the gluon polarization}

We have observed also that the present inclusive DIS data cannot
rule out the solutions with negative and changing in sign gluon
polarizations (see Fig 5a). The shape of the negative gluon
density differs from that of positive one. In all the cases the
magnitude of $\Delta G$ (the first moment of the gluon density) is
small: $|\Delta G| \leq 0.4$ and the corresponding polarized quark
densities $(\Delta u + \Delta\bar u)$ and $(\Delta d + \Delta\bar
d)$ are very close to each other. The corresponding strange sea
densities are shown in Fig. 5b. Note, however, that the
uncertainties for PDFs corresponding to the solution with $\Delta
G < 0$ are larger than those in the case of $\Delta G > 0$ (for
more details see \cite{LSS06}). In Fig. 6 the ratio $\Delta
G(x)/G(x)$ calculated for the different $\Delta G(x)$ obtained in
our analysis and using $G(x)_{\rm MRST'02}$ \cite{MRST02} for the
unpolarized gluon density, is compared to the existing direct
measurements of $\Delta G(x)/G(x)$. The error band correspond to
statistic and systematic errors of $\Delta G(x)$. The most precise
value for $\Delta G/G$, the COMPASS one, is well consistent with
any of the polarized gluon densities determined in our analysis.
One can see from Fig. 6 that in order to choose between gluons
with positive and negative polarization direct measurements of
$\Delta G(x)$ at large $x:~x > 0.3$ are needed.
\begin{figure}[t!]
\begin{center}
\begin{tabular}{cc}
\mbox{\epsfig{figure=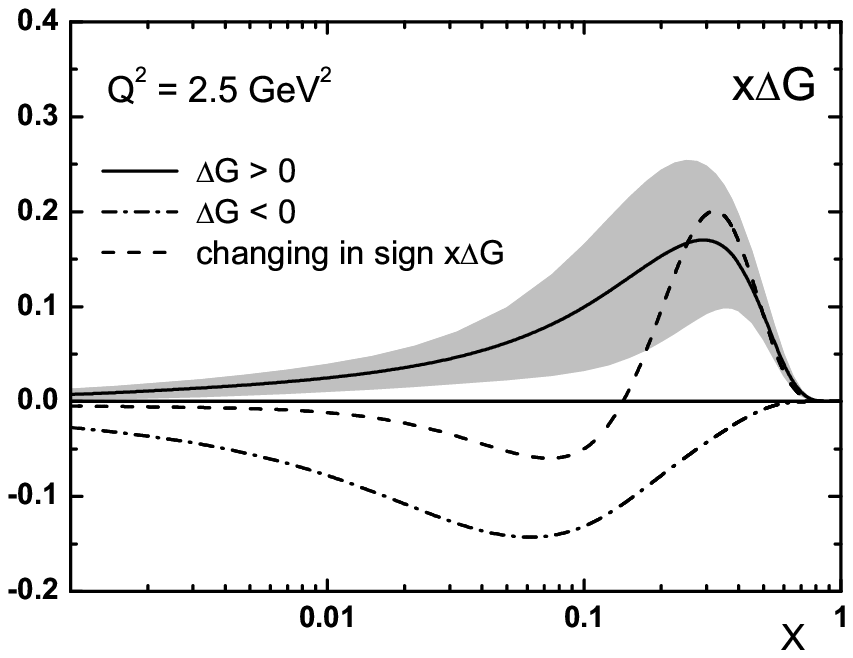,width=6.5cm,height=5cm}}&
\mbox{\epsfig{figure=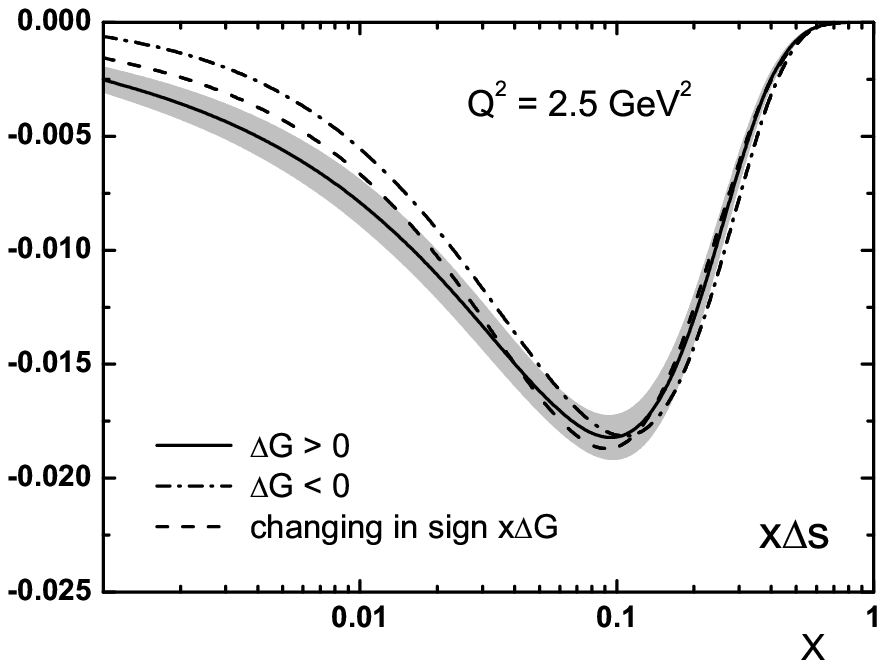,width=6.5cm,height=5cm}}\\
%{\bf(a)}& {\bf(b)}
\end{tabular}
%\centerline
\end{center}
{ {\small{\bf Figure 5.} Strange quark sea densities $x\Delta
s(x)$ corresponding to the fits with $\Delta G > 0$, $\Delta G <
0$ and changing in sign $x\Delta G$ } }
\end{figure}

\section{The proton spin sum rule and spin puzzle}

Using the values for the singlet and gluon polarizations $\Delta
\Sigma(Q^2)$ and $\Delta G(Q^2)$ at $Q^2=1~GeV^2$ obtained in our
analysis ($\rm \overline{MS}$ scheme):$\Delta\Sigma=0.207 \pm
0.039$ and $\Delta G = 0.237 \pm 0.153$ we have found the
following value for the spin of the proton at $Q^2=1~GeV^2$:

%\begin{eqnarray}
%\nonumber
\begin{equation}
S_z = {1\over 2} = {1\over 2}\Delta \Sigma(Q^2) + \Delta G(Q^2) +
L_z(Q^2) = 0.34 \pm 0.15 + L_z(Q^2). \label{SSR}
\end{equation}
%\end{eqnarray}
So, in order to satisfy the proton spin sum rule (\ref{SSR}) the
sum of the quark and gluon orbital angular momentum $L_z=L_z^q +
L_z^g$ should be different from zero and positive. Note that the
quark orbital momentum $L_z^q$ will be determined soon from the
data using the forward extrapolation of the generalized paron
densities (GPD).

Let us finally discuss the so called "spin puzzle" - the
discrepancy between the values of the singlet polarization $\Delta
\Sigma$: 0.2-0.3 in the DIS region and 0.6 at low $Q^2(Q^2 \sim
\Lambda ^2_{\rm QCD})$ (see Fig. 7a). For better understanding of
the situation it is useful to use the JET factorization scheme
\cite{JET}, in which $\Delta \Sigma(Q^2)$ does not depend on
$Q^2$. Then, in this scheme it is meaningful to directly interpret
the singlet polarization $\Delta \Sigma$ as the contribution of
the quark spins to the nucleon spin and to compare its values
obtained in the DIS and low $Q^2$ regions. The value of $\Delta
\Sigma_{\rm JET}$ obtained in our LSS'06 analysis of the DIS data
is $0.26~\pm~0.08$.

\begin{wrapfigure}[21]{R}{6.6cm}
\begin{center}
\mbox{\epsfig{figure=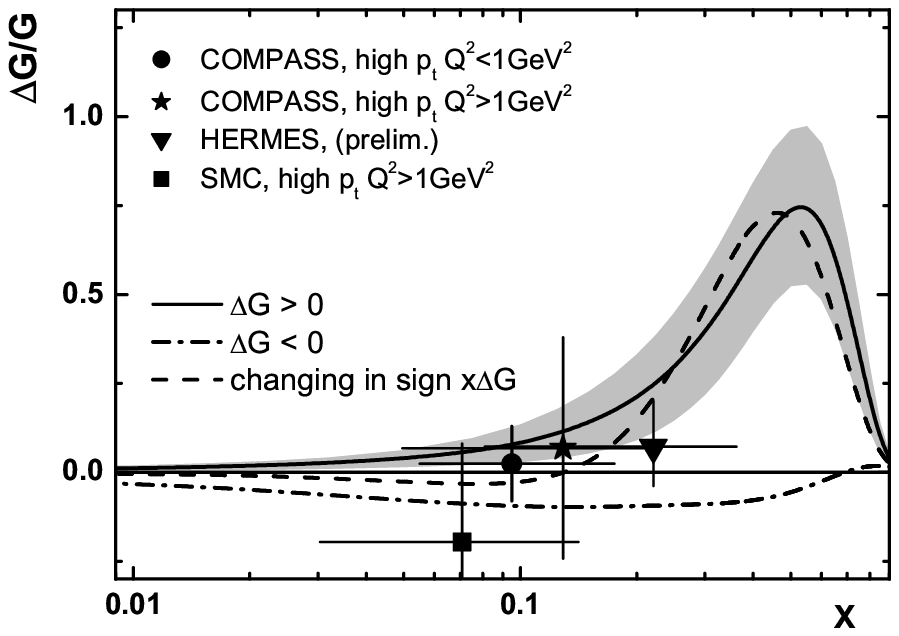,width=6.6cm,height=5cm}}
\end{center}
{\small{\bf Figure 6.} Comparison between the experimental data
and NLO($\rm \overline{MS}$) curves for the gluon polarization
$\Delta G(x)/G(x)$ at $Q^2=3~\rm GeV^2$ corresponding to $\Delta G
> 0$, $\Delta G < 0$ and an oscillating-in-sign $x\Delta G$. }
\end{wrapfigure}

On the other hand the well known value of 0.6 for $\Delta
\Sigma(Q^2 \sim \Lambda^2_{\rm QCD})=\Delta u_v + \Delta d_v +
\Delta q_{sea}$ is predicted in the relativistic constituent quark
model (CQM) \cite{RCQM}. However, this model does NOT account for
the vacuum (quark sea) polarization. It was qualitatively shown in
the instanton models \cite{spinsea,instmodel} that due to the
non-perturbative vacuum spin effects the contribution of the sea
quark polarization to $\Delta \Sigma$ is {\it negative}. So, the
value of $\Delta \Sigma$ in the non-perturbative region $(Q^2 \sim
\Lambda ^2_{\rm QCD})$ is really smaller than 0.6. Also, it was
found from a combined analysis of forward scattering
parity-violating elastic $\overrightarrow{e}p$ asymmetry data from
$\rm G^0$ and HAPPEx experiments at JLab, and elastic $\nu p$ and
$\bar{\nu}p$ scattering data from Experiment 734 at BNL, that the
strange axial form factor $G_A^S(Q^2)$, which is strongly related
with $\Delta s$~($G_A^S(Q^2=0)=\Delta s)$, is {\it negative} in
the region $0.4 < Q^2 < 1~ GeV^2$ \cite{GSA} (see Fig. 7b), {\it
i.e} there is a strong indication that the strange quark
contribution to $\Delta \Sigma$ at low $Q^2$ is {\it negative}. In
conclusion, we are very close to the solution of the so called
"spin puzzle".
\begin{figure}[b!]
\begin{center}
\begin{tabular}{cc}
\mbox{\epsfig{figure=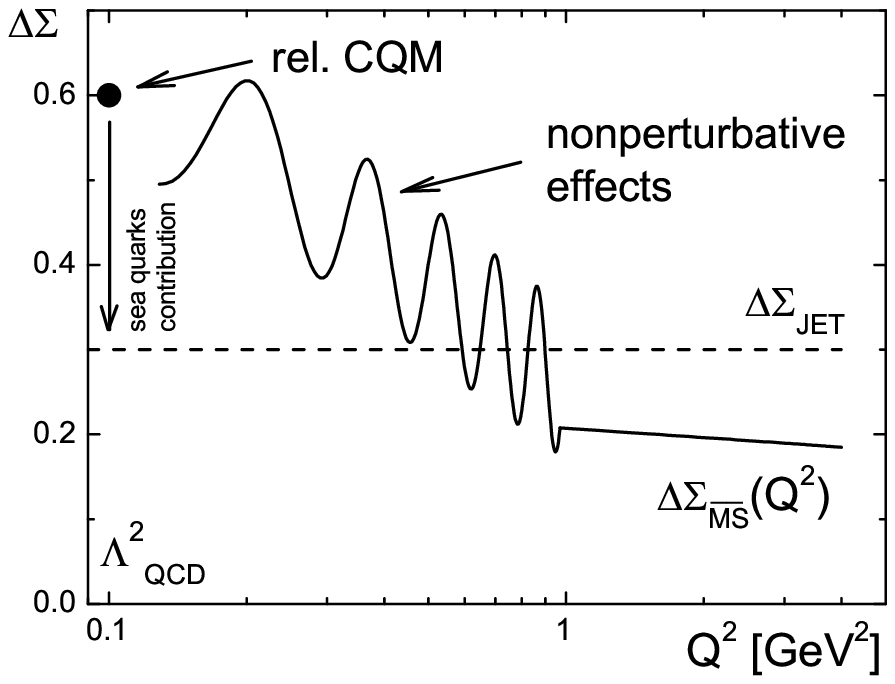,width=6.3cm,height=4.5cm}}&
\mbox{\epsfig{figure=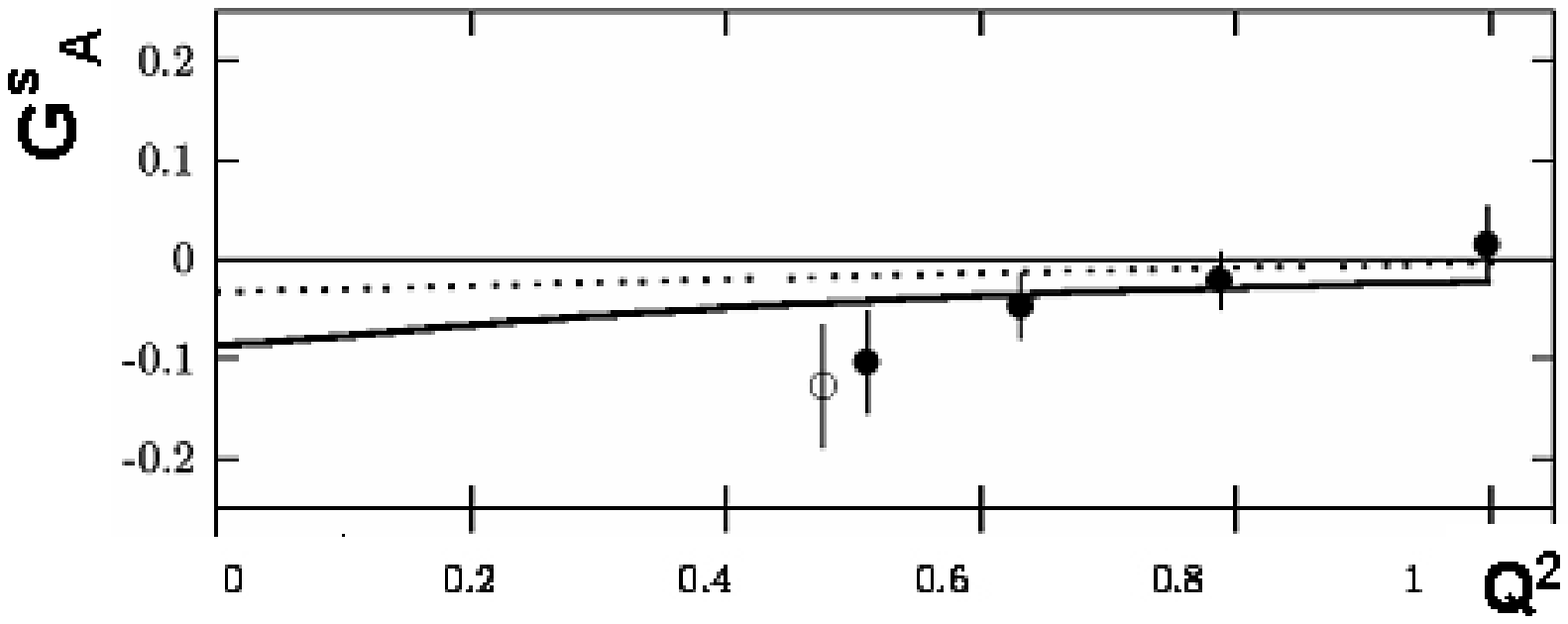,width=6.3cm,height=4.3cm}}\\
%{\bf(a)}& {\bf(b)}
\end{tabular}
\end{center}
{\small {\bf Figure 7.} A possible explanation of the nucleon's
spin puzzle (a). Results of analysis for the strange axial form
factor of the proton (b).}
\end{figure}

\section*{Conclusion}

We have studied the impact of the CLAS and latest COMPASS data on
the polarized parton densities and higher twist contributions. It
was demonstrated that the inclusion of the low $Q^2$ CLAS data in
the NLO QCD analysis of the world DIS data improves essentially
our knowledge of HT corrections to $g_1$ and does not affect the
central values of PDFs, while the large $Q^2$ COMPASS data
influence mainly the strange quark and gluon polarizations, but
practically do not change the HT corrections. The uncertainties in
the determination of polarized parton densities is significantly
reduced due to both of the data sets. These results strongly
support the QCD framework, in which the leading twist pQCD
contribution is supplemented by higher twist terms of ${\cal
O}(\Lambda^2_{\rm QCD}/Q^2)$.

Finally, one of the important messages coming from this analysis
is that it is impossible to describe the very precise CLAS data if
the HT corrections are not taken into account. Note that if the
low $Q^2$ data are not too accurate, it would be possible to
describe them using only the leading twist term in $g_1$
(logarithmic in $Q^2$), {\it i.e.} to mimic the power in $Q^2$
dependence of $g_1$ with a logarithmic one (using different forms
for the input PDFs and/or more free parameters associated with
them) which was done in the analyses of another groups before the
CLAS data were available.\\

{\bf Acknowledgments:} This research was supported by the
JINR-Bulgaria Collaborative Grant and by the RFBR Grants (No
05-01-00992, No 05-02-17748, 06-02-16215, 07-02-01046).


\begin{thebibliography}{99}

\bibitem{LSS_HT}
E. Leader, A.V. Sidorov and D.B. Stamenov, Phys. Rev. {\bf D67},
074017 (2003).
\bibitem{CLAS06}
K.V. Dharmwardane et al. (CLAS Collaboration), Phys. Lett. {\bf
B641}, 11 (2006).
\bibitem{COMPASS06}
V.Yu. Alexakhin et al. (COMPASS Collaboration), Phys. Lett. {\bf
B647}, 8 (2007).
\bibitem{NMC}
M. Arneodo et al. (NMC Collaboration), Phys. Lett. {\bf B364}, 107
(1995).
\bibitem{R1998}
K. Abe et al. (SLAC E143 Collaboration), Phys. Lett. {\bf B452},
194 (1999).
\bibitem{TMC}
A. Piccione and G. Ridolfi, Nucl. Phys. {\bf B513}, 301 (1998); \\
J.Blumlein and A. Tkabladze, Nucl. Phys. {\bf B553}, 427 (1999).
\bibitem{LSS05}
E. Leader, A.V. Sidorov and D.B. Stamenov, Phys. Rev. {\bf D73},
034023 (2006).
\bibitem{world}
J. Ashman et al. (EMC Collaboration), Phys. Lett. {\bf
B206}, 364 (1988); Nucl. Phys. {\bf B328}, 1 (1989);\\
P.L. Anthony et al. (SLAC E142 Collaboration), Phys. Rev.
{\bf D54}, 6620 (1996); \\
K. Abe et al. (SLAC/E154 Collaboration), Phys. Rev. Lett. {\bf 79}, 26 (1997);\\
B. Adeva et al. (SMC Collaboration) Phys. Rev. {\bf D58}, 112001
(1998); \\ K. Abe et al. (SLAC E143 Collaboration), Phys. Rev.
{\bf D58}, 112003 (1998); \\P.L. Anthony et al. (SLAC E155
Collaboration), Phys. Lett. {\bf B463}, 339 (1999); {\bf B493}, 19
(2000); \\ X. Zheng et al. (JLab/Hall A Collaboration), Phys. Rev.
Lett. {\bf 92}, 012004 (2004);\\A. Airapetian et al.
(HERMES Collaboration), Phys. Rev. {D\bf 71}, 012003 (2005); \\
E.S. Ageev et al. (COMPASS Collaboration), Phys. Lett. {\bf B612},
154 (2005).
\bibitem{CLAS12}
M. Amarian et al., A 12 GeV Research Proposal to Jefferson Lab,
PR12-06-109. {\it The Longitudinal Spin Structure of the Nucleon}.
\bibitem{LSS06}
E. Leader, A.V. Sidorov and D.B. Stamenov, Phys. Rev. {\bf D75},
074027 (2007).
\bibitem{MRST02}
A.D. Martin, R.G. Roberts, W.J. Stirling and R.S. Thorne, Eur.
Phys. J. {\bf C 28}, 455 (2003).
\bibitem{JET}
R. D. Carlitz, J. C. Collins and A.H. Mueller, Phys. Lett. {\bf
B214}, 229 (1988); M. Anselmino, A. V. Efremov and E. Leader,
Phys. Rep. {\bf 261}, 1 (1995); H.-Y. Cheng, Int. J. Mod. Phys.
{\bf A 11}, 5109 (1996); D. M\"{u}ller and O. V. Teryaev, Phys.
Rev. {\bf D56}, 2607 (1997).
\bibitem{RCQM}
A.W. Shreiber and A.W. Thomas, Phys. Lett. {\bf B 215}, 141
(1988); R.D. Jaffe and A. Manohar, Nucl. Phys. {\bf B 337}, 509
(1990).
\bibitem{spinsea}
S.N. Shore and G. Veneziano, Phys. Lett. {\bf B 244}, 75 (1990);
S. Forte and E.V. Shuryak, Nucl. Phys. {\bf B 357}, 153 (1991).
\bibitem{instmodel}
A.E. Dorokhov, Czech. J. Phys. {\bf 52}, c79 (2002); A.E.
Dorokhov, N.I. Kochelev and Yu.A. Zubov, Int. Journ. Mod. Phys.
{\bf A 8}, 603 (1993); A.E. Dorokhov and N.I. Kochelev, Phys.
Lett. {\bf B 304}, 167 (1993).
\bibitem{GSA}
S. Pate, AIP Conf. Proceedings {\bf 915}, 391 (2007)
(arXiv:hep-ex/0611053).

\end{thebibliography}
\end{document}